\documentclass[aps, prb, twocolumn, groupedaddress, showpacs, floatfix]{revtex4}
\usepackage{graphics}
\usepackage{dcolumn}
\usepackage{subfigure}
\usepackage{epsfig}
\usepackage{graphicx}
\usepackage{lipsum}
\begin{document}

\title{One-dimensional chain ``melting'' in incommensurate potassium}
\newcommand{\degree}{\ensuremath{^\circ}} 

\author{E. E. McBride$^{1,\ast,\dagger}$}
\author{K. A. Munro$^1$}
\author{G. W. Stinton$^1$}
\author{R.~J.~Husband$^1$}
\author{R. Briggs$^1$}
\author{H.-P. Liermann$^2$}
\author{M. I. McMahon$^{1,3}$}
\affiliation{$^1$SUPA, School of Physics and Astronomy, and Centre for Science at Extreme Conditions, The University of Edinburgh, Mayfield Road, Edinburgh, EH9 3JZ, UK}
\affiliation{$^2$Photon Science, Deutsches Elektronen-Synchrotron DESY, Notkestrasse 85, 22607, Hamburg, Germany}
\affiliation{$^3$Research Complex at Harwell, Didcot, Oxon, OX11 0FA, UK}

\begin{abstract}

Between 19 and 54 GPa, potassium has a complex composite incommensurate host-guest structure which undergoes two intraphase transitions over this pressure range. The temperature dependence of these host-guest phases is further complicated by the onset of an order-disorder transition in their guest chains. Here, we report single crystal, quasi-single crystal, and powder synchrotron X-ray diffraction measurements of this order-disorder phenomenon in incommensurate potassium to 47 GPa and 750 K. The so-called chain ``melting'' transition is clearly visible over a 22 GPa pressure range, and there are significant changes in the slope of the phase boundary which divides the ordered and disordered phases, one of which results from the intraphase transitions in the guest structure.

\end{abstract}
\date{\today}
\maketitle

\section{Introduction}
\begin{figure}[!t]
\begin{center}
\includegraphics[width=0.8\columnwidth]{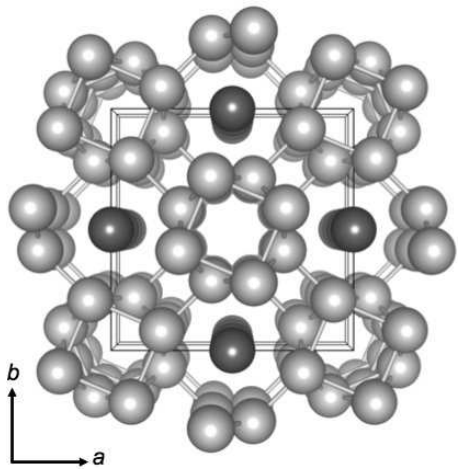}
\caption{The composite incommensurate structure of K-IIIa, as viewed down the $c$ axis. The tetragonal 16-atom host component is shown in light grey, and the guest atom chains in dark grey. A unit cell is shown by the solid lines, and the crystallographic axes are labeled.}
\label{fig:Structure}
\end{center}
\end{figure}

Of all of the complex structural forms found in the elements at high pressure, none is perhaps as intriguing as the composite incommensurate host-guest (H-G) structure. These structures, found first in barium (Ba),\cite{Nelmes1999} and then in an abundance of other Group 1, 2, 3, and 15 elements,\cite{Schwarz1999,McMahon2001,McMahon2006a,Lundegaard2009,McMahon2000,Fujihisa2013,Loa2012,Fujihisa2005,McMahon2006,
McMahon2000b,Degtyareva2004}  and which are also predicted to occur in other elements at extreme pressures,\cite{Pickard2010} comprise a tetragonal (or monoclinically-distorted) host structure with channels running along the $c$-axis. Located within these channels are 1-dimensional (1D) chains of guest atoms, which order into a number of different structures. Notably, the structure of these guest atoms is incommensurate with that of the host structure along their common $c$ axis. See Fig. \ref{fig:Structure} for an illustration of this type of structure. As a function of pressure, structural transitions to other host-guest structures with different symmetries, so-called intraphase transitions, have been observed in Ba,\cite{Nelmes1999} strontium (Sr),\cite{McMahon2000} and antimony (Sb).\cite{Degtyareva2004} Furthermore, in rubidium (Rb), near the lower pressure limit of its stability range at 300 K, between 16.2 and 16.7 GPa, the chains of guest atoms were found to lose long-range order, or ``melt'', resulting in the broadening and then near-disappearance of the Bragg reflections arising from scattering from the guest atoms.\cite{McMahon2004,Falconi2006}

The melting of the guest chains in Rb-IV has not, to date, been observed in any other elemental H-G system, although the guest atoms in the H-G phase of sodium (Na) were found to be partially disordered at room temperature over the entire pressure range studied.\cite{Lundegaard2009} Analysis of the chain melting in Rb-IV was greatly hindered by the close proximity of the disordering transition pressure to that of the reverse phase transition from Rb-IV back to Rb-III at 16.2 GPa: measurements of the Bragg peak widths needed to be collected at numerous pressures between 16.2 and 16.7 GPa, with the knowledge that reduction in pressure below 16.2 GPa would result in the loss of the Rb-IV crystal.

While studies of the melting transition in Rb-IV as a function of temperature at fixed pressure would be preferred, attempts to make such measurements were again impeded by the proximity of the Rb-IV to Rb-III transition, and by the fact that the pressure in the diamond anvil cell typically dropped below this pressure as the sample was heated.

However, in our recent studies of the melting curve of potassium (K),\cite{Narygina2011} we observed that the host-guest phase of potassium, K-III, also exhibits an order-disorder transition of the guest chains between 350 and 420 K at approximately 21 GPa.\cite{McBride2012} Our first observation of this phenomenon in quasi-single crystal studies showed that the transition was completely reversible on temperature decrease, and occurred at pressures and temperatures that were distant from either the K-II to K-III transition, or the complete melting of the sample, thereby aiding studies of the disordering phenomenon.

In this paper we describe detailed studies of this order-disorder transition in K-III near 20 GPa using a high-quality single crystal of K-III, and a study of the pressure dependence of the order-disorder transition to 47 GPa and 750 K.

\section{Experimental Details}\label{experiment}

Diffraction data were collected from single-crystals, quasi-single crystals and polycrystalline samples of K-III at the Diamond Light Source (DLS), PETRA III, and the European Synchrotron Radiation Facility (ESRF) synchrotrons, on beamlines I15, P02.2, and ID09a, respectively. At the DLS and the ESRF, the incident X-ray wavelength was $\sim$0.41 \AA\, and the diameter of the incident X-ray beams were 20 $\mu$m and 15 $\mu$m, respectively. At PETRA III, the incident wavelength was 0.29 \AA\ and the incident beam diameter was 2 $\mu$m. The data were collected on Mar345 (DLS and PETRA III), Mar555 (ESRF) and Perkin-Elmer XRD 1621 (PETRA III) detectors, which were placed $\sim$400 mm (DLS, ESRF, Perkin-Elmer at PETRA III) and $\sim$450 mm (Mar345 at PETRA III) from the sample. The majority of the data used to establish the phase diagram to 47 GPa were collected on ID09a using gas-membrane driven diamond anvil cells.\cite{Jenei2013}

All experiments were performed using high-purity (99.95+\%) potassium obtained from the Sigma-Aldrich Chemical Company. To prevent oxidation of the samples, the K was loaded into diamond-anvil pressure cells without a pressure-transmitting medium in a dry oxygen-free atmosphere ($<$1 ppm O$_2$ and $<$1 ppm H$_2$O). While the 99.95+\% stated purity of the sample refers to the metal content, such samples may contain other non-metallic contaminants such as oxygen and hydrogen. However, in our previous study of potassium using both as-purchased and distilled samples,\cite{McMahon2006a} we observed no difference in their high-pressure behavior. Diffraction patterns of the low-pressure \textit{bcc} and \textit{fcc} phases showed no discernible contaminant peaks.

For the high-temperature studies, the sample temperature was controlled using external resistive heating rings (Watlow Ltd.) up to temperatures of 750 K. The sample temperature was measured using a K-type thermocouple, which was attached to one of the diamond anvils, close to the sample. All samples were loaded with small pieces of ruby or a few grains of tantalum (Ta) powder in order to determine the pressure using either the ruby fluorescence technique,\cite{Mao1986} or the equation of state of Ta.\cite{Dewaele2004} To account for the thermal expansion of the Ta lattice in the high-temperature studies, a temperature correction was applied to the Ta equation of state, following Cohen \& G\"ulerson.\cite{Cohen2001}


\section{Results}

\subsection{The Structure of K-III at 20.3 GPa and Room Temperature}\label{SingleXtal}

Our initial observation of the order-disorder transition in a quasi-single crystal of K-III revealed that, as in Rb-IV, the Bragg reflections from the guest component of the structure are elongated perpendicular to the crystallographic $c$-axis as a consequence of the decrease in the coherent scattering between neighbouring chains of guest atoms.\cite{McBride2012} This suggested that in order to study the order-disorder transition in K-III in more detail, a single-crystal sample would be required.

\begin{figure}[!t]
\begin{center}
\includegraphics[width=0.9\columnwidth]{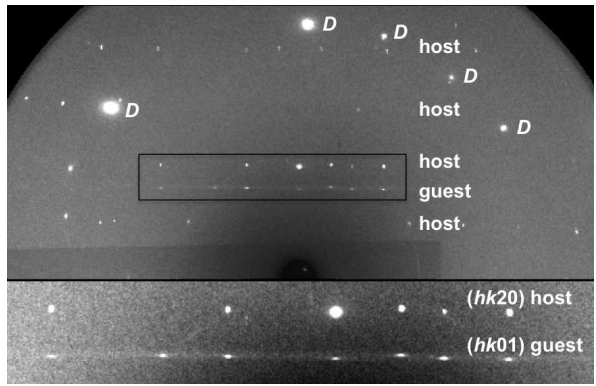}
\caption{Composite 2D diffraction image from the high quality single crystal of K-IIIa at 20.3 GPa and 295 K. This composite image is formed from the summation of several 1$^\circ$ contiguous images obtained during the collection of a single-crystal data set. The layer lines of host and guest reflections are labeled, and reflections from the diamond anvils are labeled with `\textit{D}'. The lower inset shows an enlarged view of the ($hk$20) host and ($hk$01) guest reflections, highlighting the distinct differences in their shape. The host reflections are sharp Bragg spots, as expected from a 3D crystal, while the guest reflections are disc-shaped, elongated perpendicular to the 
crystallographic $c$ axis. The ($hk$01) guest reflections are located on a sheet of diffuse scattering (seen as a line in this projection) that arises from the 1D nature of the guest chains.}
\label{fig:Composite}
\end{center}
\end{figure}

The growth of such a crystal was greatly aided by knowledge of the melt curve,\cite{Narygina2011} and melting and high-temperature annealing at $\sim$21 GPa resulted in the growth of a small, very high-quality single crystal of K-III at 20.3 GPa. Diffraction data from the crystal were collected on beamline P02.2 at PETRA III in a sequence of contiguous 1$^{\circ}$ oscillations over a total scan range of $\pm$25$^{\circ}$ around the vertical axis. Data were collected with the $c$-axis of the sample both in the horizontal plane, and at approximately 45$^{\circ}$ to the horizontal. An exposure time of 1 second per frame was used, with the incident X-ray beam suitably attenuated so as to ensure that the strongest reflections were not saturated. The frames were then processed as described by McMahon \textit{et al.}\cite{McMahon2013} to yield accurate Bragg intensities for both host and guest reflections. An example of the quality of the diffraction data obtained is given in Fig. \ref{fig:Composite}. Even in the data collected at 295 K one can see that while the Bragg peaks from the host component of the structure are small round spots, the Bragg peaks from the guest component are markedly different, forming discs elongated perpendicular to the $c$-axis.

Prior to making a high-temperature study of the order-disorder transition in K-III (see Section \ref{powderstudy}), we used the single crystal data to resolve some of the outstanding structural uncertainties remaining from previous studies. K-III has the 4D superspace group $I4/mcm(00q_3)000s$. However, we have previously noted that at $\sim$27 GPa, the ($1010$) reflection from the host component of the structure appears in powder-diffraction profiles collected from this phase, suggesting a loss of the $c$-glide symmetry in the host substructure.\cite{Lundegaard2013} Unfortunately, the other ($h0l0$) $l$ = odd reflections that would have confirmed the absence of $c$-glide symmetry were overlapped by other reflections in the powder patterns. Such overlapping reflections would not occur in single-crystal data. Analysis of the data collected from the single crystal of K-III at 20.3 GPa revealed that the systematically absent reflections were completely consistent with the host component of the structure having space group $I4/mcm$. In particular, there was no evidence of any ($h0l$0) with $l$ = odd reflections that would indicate an absence of the $c$-glide symmetry. The 4D superspace group is therefore confirmed as $I4/mcm(00q_3)000s$, at least at 20.3 GPa.

We also used the single-crystal data to determine whether any of the weak ($hklm$) modulation reflections (with $l$ and $m$ $\neq$ 0) that result from the interaction between the host and the guest components of the structure, were present. We have observed such modulation reflections previously in Bi and Sb.\cite{McMahon2000,Degtyareva2004,McMahon2007}. No such modulation reflections were observed, suggesting that the host-guest interactions in K-III are extremely weak even at room temperature.

Refinements of the single-crystal data were performed using {\sc jana\oldstylenums{2006}}.\cite{Jana} and its quality was sufficient to obtain the anisotropic atomic displacement parameters (ADPs) of the host and guest atoms. The ADPs for the host atoms are $U_{11}$ = 0.028(1) \AA$^2$, $U_{22}$ = 0.024(1) \AA$^2$, $U_{33}$ = 0.033(1) \AA$^2$, and $U_{23}$ = -0.001(1) \AA$^2$, suggesting an almost isotropic motion of the atom. However, the ADPs of the guest atom -- $U_{11}$ = 0.029(3) \AA$^2$, $U_{22}$ = 0.024(3) \AA$^2$, and $U_{33}$ = 0.170(3) \AA$^2$ -- demonstrate that its atomic motion is strikingly anisotropic: while the RMS displacement of the guest atoms perpendicular to the chain direction is $\sim$0.17 \AA, very similar to that of the host atoms, the RMS displacement along the $c$-axis is $\sim$0.41 \AA, some 2.4 times larger. This increased thermal motion along $c$ is very evident in the diffraction data -- while ($hkl0$) host reflections with $l > 1$ are easily discernible in the diffraction patterns, ($hk0m$) guest reflections with $m > 1$ are not. A simple comparison of the average displacement of the guest atoms along the chains and the spacing of the guest atoms in the chains ($c_G$ = 2.989(3) \AA) shows that the RMS displacement is nearly 14 \% of the spacing -- very close to the Lindemann criterion for melting.\cite{Lindemann1910}

\begin{figure}[!t]
\begin{center}
\includegraphics[width=0.8\columnwidth]{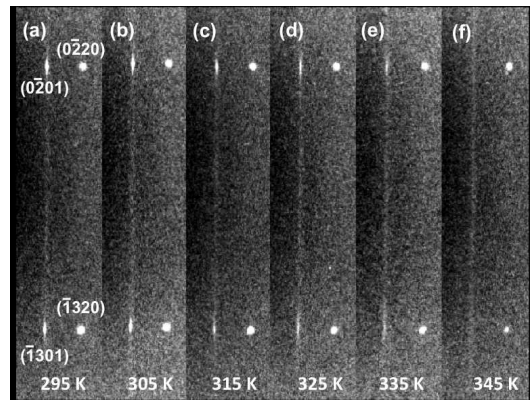}
\caption{Enlargements of 2D diffraction images collected from the single crystal of K-IIIa at 20.3 GPa, highlighting the ($0\bar{2}20$) and ($\bar{1}320$) host reflections, and the ($0\bar{2}01$) and ($\bar{1}301$) guest reflections. The images were obtained using a 1$^\circ$ oscillation of the crystal. Panels (a) to (f) show the changes in the diffraction peaks as the sample temperature was increased from 295 K to 345 K in 10 K steps. The host reflections remain unchanged with increasing temperature, whereas the guest reflections first weaken and broaden (305 and 315 K), then the Bragg scattering disappears and only discs of diffuse scattering remain (325 and 335 K), before all structure in the scattering disappears, and only a uniform sheet of diffuse scattering remains (345 K).}
\label{fig:StackChainMelt}
\end{center}
\end{figure} 

\subsection{High Temperature Studies of K-III at 20.3~GPa}\label{SingleXtal2}

Following the detailed study of the K-III structure at 295 K, the sample temperature was increased in 10 K steps at 20.3 GPa. Rotation images were collected at each temperature in order to follow changes in the shape of the Bragg reflections from the host and guest components of the structure, and complete single-crystal data sets were collected at every temperature step. While there were no discernible changes in the reflections from the host substructure on temperature increase, changes in the shape of the Bragg reflections from the guest substructure were clearly evident, as illustrated in Fig. \ref{fig:StackChainMelt}. As the temperature was increased, the guest reflections elongated further perpendicular to the $c$-axis, and decreased in intensity. At temperatures of 325 K and above, only a disc of diffuse scattering was evident, suggesting a loss of the long-range ordering in the guest structure, and at 335 K the scattering became an essentially uniform sheet of diffuse scattering, indicating a loss of even short-range ordering.

The interchain correlation length can be quantified from the broadening of the guest reflections.\cite{McMahon2004} The width of the ($0\bar{2}01$) and ($\bar{1}301$) guest reflections perpendicular to the chain direction were obtained as a function of temperature by integrating them into a 1D profile along the direction of the line of diffuse scattering. Integrations were performed using the \textsc{fit\oldstylenums{2}d} software package.\cite{Fit2d} The additional width over that measured from the ($0\bar{2}20$) and ($\bar{1}320$) host reflections at 295 K, $\Delta_{FWHM}$, can be used to estimate the correlation length via $2\pi/\Delta_{FWHM}$. The dimension of the host reflection along the relevant direction at 295 K corresponds to a $2\pi/\Delta_{FWHM}$ value of $\sim$ 300 \AA\, which is thus the effective resolution limit of these measurements. The peak-width of the host reflections were found to have negligible variation over the temperature range studied.

\begin{figure}[!b]
\begin{center}
\includegraphics[width=1.0\columnwidth]{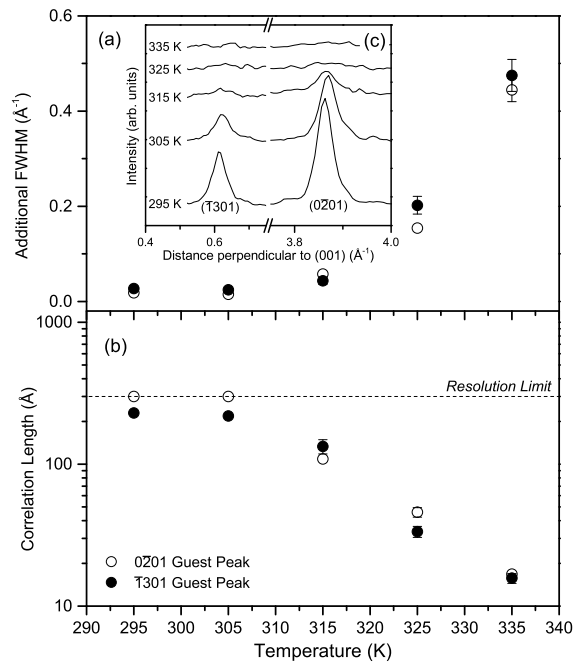}
\caption{Panel (a) shows the additional peak-width of the ($0\bar{2}01$) and ($\bar{1}301$) guest reflections (plotted with unfilled and filled symbols, respectively) over that of the host reflections, perpendicular to the (001) direction at 20.3 GPa as a function of temperature. The additional peak width was determined by subtracting the peak-width of the host peaks, which were found to have negligible variation in width with temperature. Panel (b) shows the effect of temperature on the interchain correlation length of the guest chains. The resolution of this measurement was determined from the peak width of the host reflection, and found to be 300 \AA. There is an exponential reduction in correlation length above 315 K. In panel (a) and (b), error bars which are smaller than the symbol size and have been omitted for clarity. Inset (c) shows 1D diffraction profiles of the
$(0\bar{2}01$) and ($\bar{1}301$) guest reflections integrated along a line perpendicular to (001). One may see that as temperature increases the guest reflections reduce in intensity before disappearing entirely above 335 K.}
\label{fig:CorrLength}
\end{center}
\end{figure}

The temperature dependence of the interchain correlation length is shown in Fig. \ref{fig:CorrLength}. The guest reflections broaden rapidly above 305 K, and the correlation length drops exponentially. At 325 K, where the absence of any Bragg scattering shows that the guest structure is no longer long-range ordered, the interchain correlation length has dropped to only $\sim$30 \AA\ or 4 interchain spacings, very similar to that found in the H-G phase of Na at 147 GPa and room temperature,\cite{Lundegaard2009} and in the H-G phase of Rb at 16.2 GPa.\cite{McMahon2004} At 335 K, the highest temperature at which the peak widths could still be determined with any certainty, the interchain correlation length is only 15 \AA\ or 2 interchain spacings. At 345 K, no structure is visible in the diffuse scattering from the chains, indicating a loss of even any short range order, and revealing that the guest chains are a true 1D liquid within the crystalline host structure.\cite{Emery1978,Spal1980,Falconi2006}

In the 1D harmonic liquid model,\cite{Emery1978,Spal1980} the widths of the resulting sheets of diffuse scattering are proportional to the square of the order of the sheet,\cite{Heilmann1979} and thus if the guest chains in K-III at 345 K are truly a 1D liquid, then the width of the second-order diffuse sheet should be four times that of the first. Several long exposures were taken to see if we could confirm this relationship in the widths, but without success; the scattering in the second sheet was too weak to determine its width, even when summing multiple long-exposure images.


\begin{figure*}[ptb]
\includegraphics[width=\textwidth]{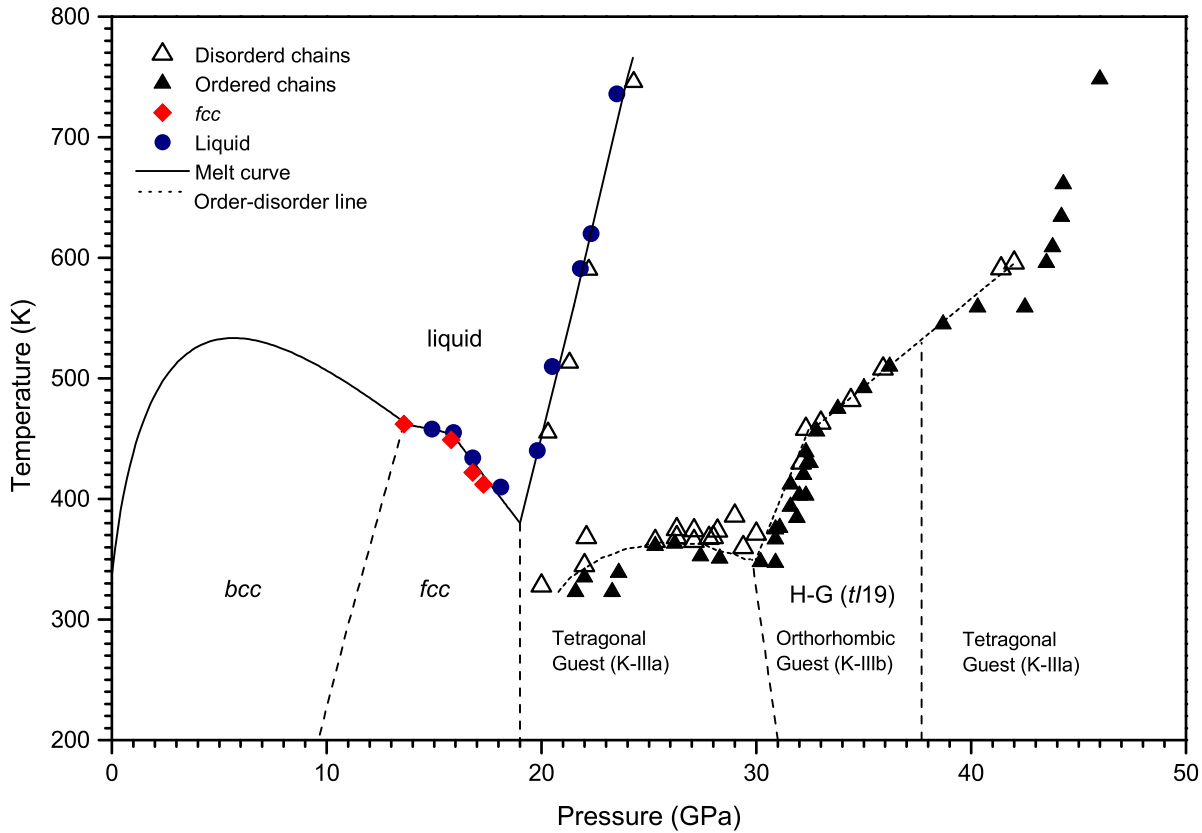}
\caption{ The phase diagram of potassium to 50 GPa and 800 K (color online). The melting curve is shown by the solid line and is a combination of the study by Narygina \textit{et al.}\cite{Narygina2011} and the additional data obtained in this study. The P-T conditions at which the \textit{fcc} and liquid phases were observed in the current study are shown by the filled (red) diamonds and the filled (blue) circles, respectively. The P-T conditions at which the incommensurate host-guest phases were observed are shown by the open and closed (black) triangles. Closed (black) triangles show where ordered guest chains were observed, and open triangles show where  disordered guest chains were observed. The dot-dash line through these points is the best-fitting order-disorder transition line, and is drawn as a guide to the eye. The observation of the K-IIIa $\rightarrow$ K-IIIb transition at 29.6 GPa and 340 K, results in a negative gradient of the transition line of $\sim$70 K/GPa,  whereas the re-entrant transition back to K-IIIa at 38 GPa and 530 K allows the almost vertical transition line to be determined -- transition lines are shown with dashed lines. No disordering of the chains was observed above 42 GPa and 596 K.} 
\label{fig:SimonMelt}
\end{figure*}

\subsection{Powder \& Quasi-single Crystal Studies of Chain Melting to 47 GPa}\label{powderstudy}

While single-crystal diffraction studies provided a wealth of information on the order-disorder phenomenon at $\sim$20~GPa, high-quality single-crystals of the H-G phase at higher pressures were extremely difficult to grow from the melt due to the steep slope of the melting curve above that pressure.\cite{Narygina2011} To investigate the order-disorder transition temperature in the guest chains at higher pressures, both polycrystalline and quasi-single crystal samples were therefore utilized. Data were collected at all three synchrotron facilities described in Section \ref{experiment}. The 2D diffraction images were integrated azimuthally using the \textsc{fit\oldstylenums{2}d} software package,\cite{Fit2d} and the resulting integrated profiles were analyzed using the Le Bail method and the \textsc{jana\oldstylenums{2006}} software package.\cite{Jana}

In addition to investigating the order-disorder transition, the same samples also allowed us to extend our previous measurements of the melting curve of K to 24 GPa and 750~K. Prior to extending the melting curve at the highest pressures, additional melting points were obtained in the pressure range 14-17 GPa, which, in combination with our previous melting data, demonstrated that following the \textit{bcc}-\textit{fcc}-liquid triple point at 13.6 GPa and 462 K, the melting temperature decreases slightly, rather than plateauing as reported previously.\cite{Narygina2011}

The revised melting curve of K is shown in Fig. \ref{fig:SimonMelt}, and the slope of the melting curve above the {\it fcc}-H-G-liquid triple point at 19 GPa and 390 K is a constant 74 K/GPa. Up to 24 GPa, we see no evidence of the slope of the melting curve becoming less steep, which might have suggested a further melting maximum above 24 GPa, and we see no evidence of melting below 750~K at pressures between 24 and 47 GPa.

\begin{figure*}[ptb]
\includegraphics*[width=\textwidth]{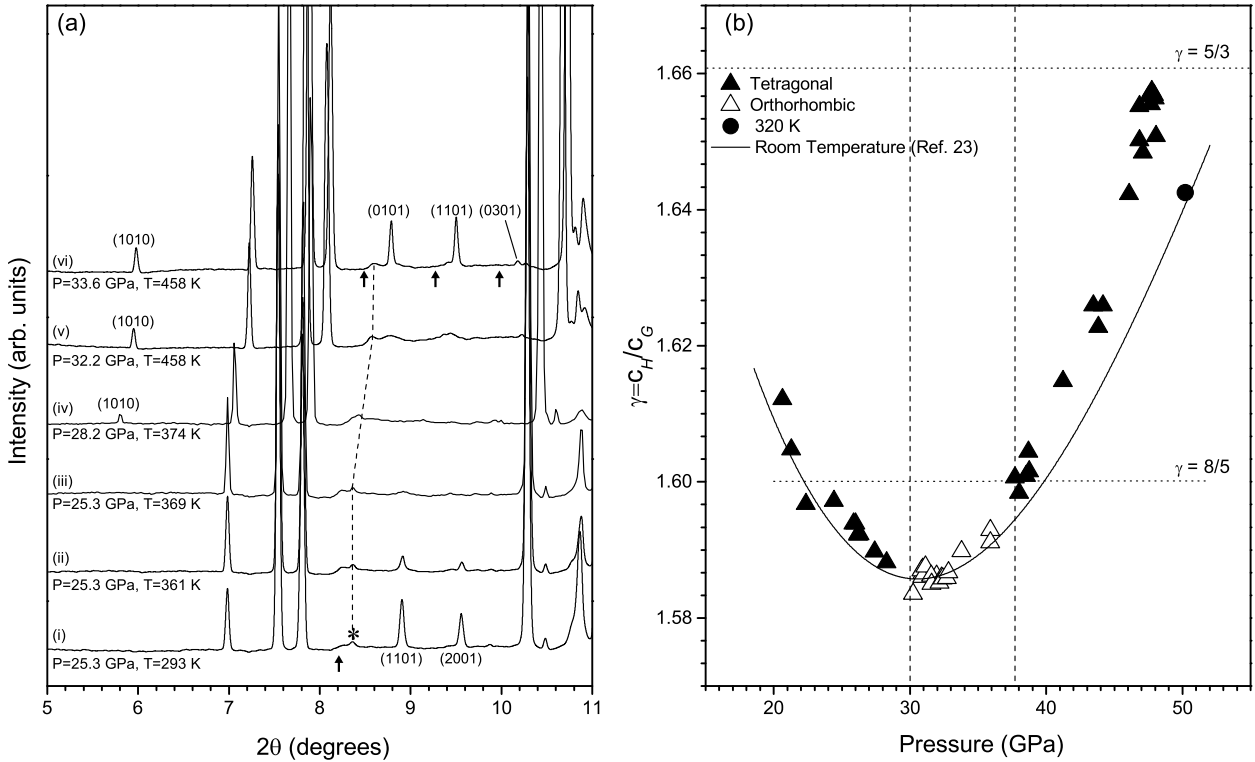}
\caption{(a) 1D integrated diffraction profiles from potassium as a function of pressure and temperature, showing the disappearance of the guest reflections in tetragonal K-IIIa (marked with their indices (1101) and (2001) in profile (i)), on temperature increase at 25.3 GPa, profiles (i) to (iii). The arrow below profile (i) shows the calculated position of the (0001) guest reflection, which is absent due to the large degree of preferred orientation in the sample. The asterisk marks a weak contaminant peak. On further increases in P and T, the (1010) host reflection appears, which signals the loss of $c$-glide symmetry, profile (iv), and the guest reflections reappear (marked with their indices (0101), (1101) and (0301)) as the structure orders into orthorhombic K-IIIb, profiles (v) and (vi). The arrows beneath profile (vi) mark the calculated positions of the guest reflections if the guest structure was tetragonal. (b) The pressure dependence of the $c_{H}$/$c_{G}$ ratio at high temperature, along the order-disorder line. Closed triangles indicate where the guest structure is tetragonal in K-IIIa, and open triangles indicate where the guest structure is orthorhombic in K-IIIb. Dotted lines indicate commensurate values of $c_{H}$/$c_{G}$. The solid curve is a fit to the room temperature data of the pressure dependence of the  $c_{H}$/$c_{G}$ ratio by Lundegaard {\it et al.},\cite{Lundegaard2013} and the closed circle indicates a high-pressure data point from the current study, close to room temperature, in good agreement with the data from Lundegaard {\it et al.} The dashed vertical lines are the room temperature intraphase transition pressures.\cite{Lundegaard2013}}
\label{fig:TetOrtho1D}
\end{figure*}

For the polycrystalline and quasi-single crystal samples, the onset of the order-disorder transition was identified by the abrupt appearance or disappearance of Bragg scattering from the guest component of the structure. Importantly, at $\sim$20 GPa, excellent agreement was found between the transition temperature determined from powder-like samples and the single-crystal sample discussed in Section \ref{SingleXtal2}. By making small increases in pressure and temperature, as illustrated in Fig. \ref{fig:TetOrtho1D}(a), the order-disorder transition line was measured up to 47~GPa using four polycrystalline samples, and is shown in Fig. \ref{fig:SimonMelt}. 

Beginning at 20~GPa, the order-disorder phase boundary has a positive slope of $\sim$10~K/GPa, reaching a maximum at $\sim$27~GPa and 365 K, before reducing slightly to 350 K at $\sim$29 GPa. We note that a similar phenomenon has been observed in the compound Hg$_{3-\delta}$AsF$_6$, where the temperature at which the Hg-chains lose long-range order also increases with pressure, but at the much higher rate of 120~K/GPa.\cite{Axe1983} Between 25.3 GPa and 28.2 GPa, the emergence of the ($1010$) host reflection, as illustrated in profiles (iii) and (iv) of Fig. \ref{fig:TetOrtho1D} (a), marks the loss of $c$-glide symmetry in the host structure, as observed previously at 26.5 GPa at 295 K. It is notable that the loss of the $c$-glide symmetry of the host occurs in both the ordered and disordered phases (Fig. \ref{fig:TetOrtho1D} (a), profile (iv)), and that the loss of symmetry occurs at a similar pressure to the slight maximum observed in the order-disorder temperature. At $\sim$29~GPa there is a sharp change in slope of the order-disorder phase boundary, which then increases at a rate of 42 K/GPa. This change in gradient is accompanied by a phase transition in the guest substructure in the ordered phase: below 29 GPa, the guest has tetragonal symmetry and the ordered structure is that of K-IIIa, while above that pressure the guest is orthorhombic, and the ordered structure is that of K-IIIb, as described by Lundegaard \textit{et al.} at room temperature.\cite{Lundegaard2013} At 295 K, the K-IIIa $\rightarrow$ K-IIIb transition takes place with no volume change. This is not the case at 350 K, where there is a volume discontinuity of $\sim$2\%, coming from small discontinuities in both the $a$ and $c$ lattice parameter of the host structure, as shown in Fig. \ref{fig:Lattice}.  There appears to be no accompanying change in $c$ lattice parameter of the guest structure at the same pressure, although the absence of any measurements of $c_G$ in the disordered phase in this pressure range makes a definitive statement difficult.

The slope of the order-disorder transition line remains constant until 32.3~GPa where it reduces to $\sim$14~K/GPa. There are no changes in either the host or guest structure that accompany this change in slope, but this pressure does coincide with a change in compressibility of $c_{H}$ -- as shown in Fig. \ref{fig:Lattice}. At 37.7~GPa the ordered guest substructure transforms \textit{back} to the tetragonal structure of K-IIIa it assumed below 29~GPa, while at $\sim$42~GPa the host regains its {\it c}-glide symmetry, as signified by the disappearance of the (1010) host reflection. Above 42~GPa, therefore, the host-guest structure is that of K-IIIa, the same as it adopted below 29~GPa. However, neither the re-entrant phase transition in the guest, nor the reappearance of {\it c}-glide symmetry in the host, appears to have any influence on the order-disorder transition temperature -- between 32~GPa and 42~GPa the slope of the transition line is constant at $\sim$14 K/GPa. Finally, at 42.5~GPa, there is a steep increase in the slope of the phase line, and indeed above this pressure we were unable to disorder the guest chains -- only the ordered K-IIIa structure was observed up to 47 GPa and 750 K.

The evolution of the incommensurate wavevector $\gamma$ = $c_{H}$/$c_{G}$ along the order-disorder phase boundary is shown in Fig. \ref{fig:TetOrtho1D} (b), and is very similar to that observed at room temperature.\cite{Lundegaard2013} $\gamma$ initially decreases with pressure, and passes through the commensurate value of $\frac{8}{5}$ at 22 GPa before reaching a minimum value of 1.585 at $\sim$30 GPa, very close to the transition to orthorhombic K-IIIb. The small discontinuity in c$_H$ at 29 GPa (see Fig. \ref{fig:Lattice}) does not result in any obvious discontinuity in $\gamma$ at the same pressure. Above 30 GPa, $\gamma$ increases, probably as a result of the change in compressibility of $c_H$ at this pressure, passing once more through the commensurate value of $\frac{8}{5}$ at 38 GPa at the transition back to tetragonal K-IIIa, before increasing more rapidly. At 47 GPa and 750 K, the highest pressure and temperature reached in this study, $\gamma$ approaches a second commensurate value of $\frac{5}{3}$. A comparison of the P-T dependence of $\gamma$ along the order-disorder phase boundary with that obtained at 295 K shows that temperature has little discernible effect up to 35 GPa, where the order-disorder temperature is $\sim$ 500 K, suggesting that the $c$-axis thermal expansions of the host and guest are the same up to this pressure. However, above 38 GPa in re-entrant K-IIIa, where the order-disorder temperature is above 530 K, $\gamma$ increases with increasing temperature, suggesting that the thermal expansion of the host structure along $c$ is somewhat greater than that of the guest.

\section{Discussion}

The pressure dependence of the order-disorder transition temperature in K-III reveals a number of striking changes in the slope of the phase line. A full interpretation for these variations in terms of changes in the density and entropy at the transition is hampered by a lack of knowledge of the structure, and hence density, of the disordered phase. However, a number of the changes in slope can be associated with changes in the structure of the host component. The general increase in the disordering transition temperature with increasing pressures suggests a strengthening of interaction between the host and guest components as the interchain distance decreases, as also reported by Axe {\it et al.} in their high-pressure study of Hg$_{3-\delta}$AsF$_6$.\cite{Axe1983} As this interaction is manifested in the K-III diffraction data by the presence of the ($hklm$) modulation reflections (with $l$ and $m$ $\neq$ 0), we would then expect these reflections to become more intense with increasing pressure. Unfortunately, no modulation reflections were visible in any of the powder profiles (including those obtained previously at room temperature\cite{Lundegaard2013}), and high-quality single crystals will be required to determine whether they do indeed become more intense at high pressures.

\begin{figure}[!b]
\begin{center}
\includegraphics[width=1.0\columnwidth]{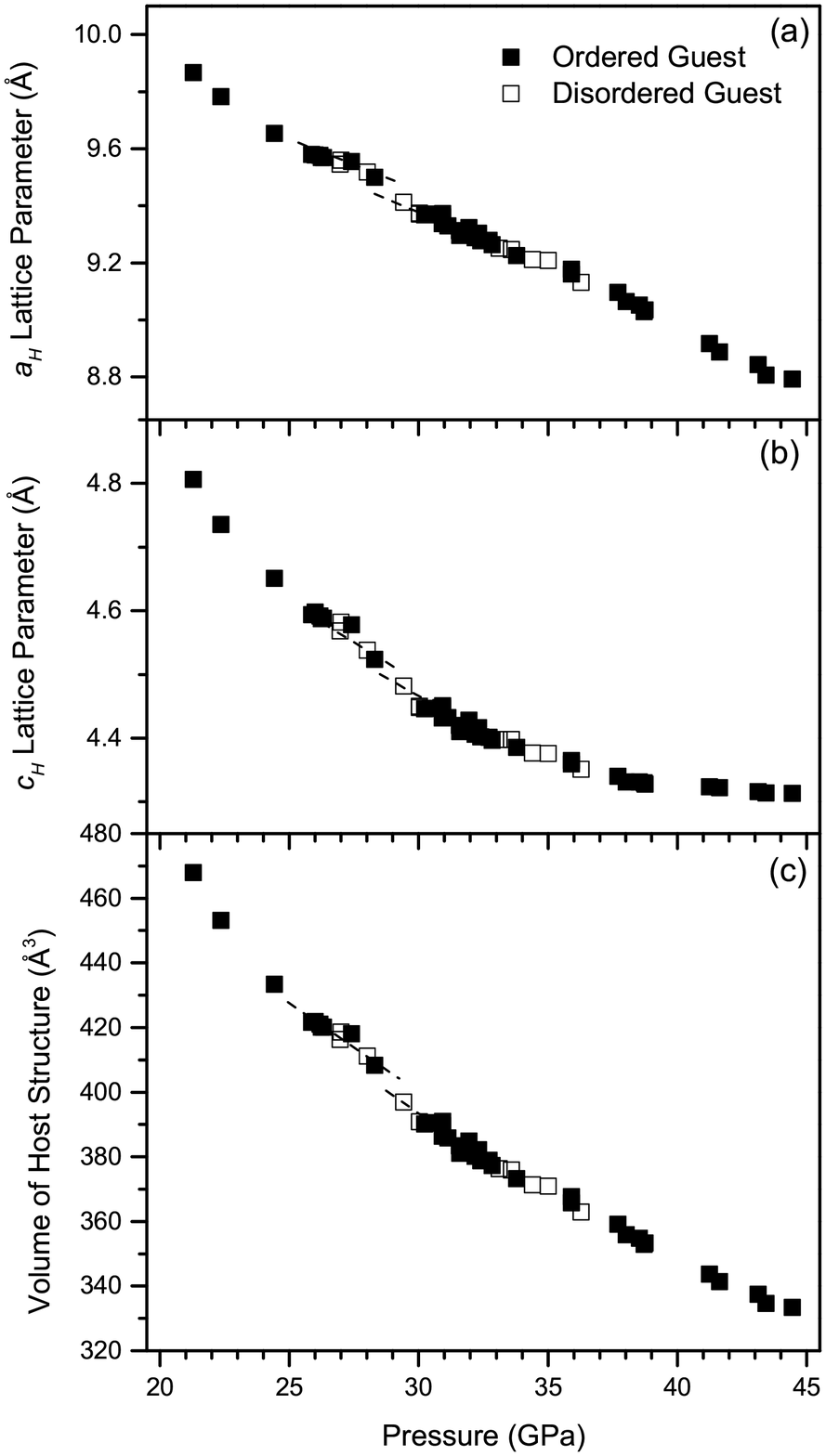}
\caption{The pressure-dependence of the (a) $a_H$,  (b) $c_H$, and (c) volume of the host unit cell along the order-disorder phase line. $a_H$ and $c_H$ can be determined in both the ordered and disordered phases, while $c_G$ can be determined only in the ordered phase. At $\sim$29 GPa, there is a volume discontinuity of $\sim$2\% in the host structure, due largely to a discontinuity in the $a_H$ lattice parameter. Above 29 GPa the $a_H$ lattice parameter decreases steadily with increasing pressure, whereas there is a distinct change in the compressibility of the $c_H$ lattice parameter near $\sim$32 GPa. This is not accompanied by a change in structure, yet does coincide with a change in gradient of the order-disorder transition line.}
\label{fig:Lattice}
\end{center}
\end{figure}

The changes in the interaction between the host and guest with pressure might also be estimated from their interatomic separation via the relative dimensions of the diameters of the host channels, which can be determined from the atomic coordinates of the host structure, and the size of the guest atoms, which can be approximated by the $c_G$ lattice parameter. If the diameters of the host channels compress more rapidly than $c_G$, then we might expect the interaction between the host and the guest to increase, due to their increased proximity, and the order-disorder temperature to increase. Unfortunately, while $c_G$ is straightforward to determine, the variable atomic coordinates of the host atom could not be determined with any confidence, particularly at the higher pressures, due to the quasi-single crystal nature of the samples, which prevented the determination of accurate Bragg intensities. However, we would note that the disappearance of the $c$-glide symmetry in the host structure between 25.3 and 28.3 GPa (Fig. \ref{fig:TetOrtho1D} (a), profiles (iii) and (iv)), which would certainly result in some change in the geometry of the channels in the host, seems to have the effect of {\it decreasing} the ordering temperature -- although the effect is small.

In contrast to the room temperature study of Lundegaard {\it et al.},\cite{Lundegaard2013} which observed no volume discontinuity at the K-IIIa to K-IIIb transition, we observe a $\sim$2\% decrease in the volume at the same transition at $\sim$29 GPa and $\sim$350 K. A volume discontinuity also occurs at the same pressure in the disordered phase (at least in the host structure - the density of the guest component in the disordered phase cannot be determined with any accuracy), making it difficult to equate the change in slope of the order-disorder phase line at 29 GPa solely with the density change. Rather, the discontinuity in $a_H$ at $\sim$29 GPa may lead to a change in geometry of guest-atom channels, increasing the host-guest interaction and thus the order-disorder temperature. As said, the quasi-single crystal nature of the samples prevented a determination the exact geometry of the host structure, and further insight into the pressure-dependence of the disordering temperature will require input from \textit{ab initio} electronic structure calculations, including calculations to determine the entropy change that occurs on disordering of the chains. Such calculations are currently ongoing in K and Rb, and will be the subject of a subsequent study. 


The presence of T-induced disorder in the guest-atom chains of incommensurate K-III reported in this study, and the previously-reported P-induced order-disorder transition in Rb at room temperature,\cite{McMahon2004} has implications for the behavior of the incommensurate host-guest structure in Na observed between 125 and 200 GPa.\cite{Lundegaard2009} In Na, the guest chains adopt a monoclinic structure, and show a high degree of disorder from 125 GPa to at least 155 GPa, the highest pressure at which their structure has been determined.\cite{Lundegaard2009} We previously attributed this unusual behaviour to the proximity of the Na melt curve to room temperature even at 150 GPa, the current results suggest another possibility -- that there is an order-disorder transition in the guest chains below 295 K, the temperature of which is not strongly pressure dependent. Low-temperature studies of incommensurate Na at $\sim$130 GPa are required to determine whether such a transition exists.

Finally, we note that at all pressures above 42 GPa, the H-G phase of K was observed with only ordered guest chains, up to the highest temperature of 750 K reached in this study. This suggests that either the slope of the order-disorder phase line increases very sharply at 42 GPa, with no apparent change in the structure (see Fig. \ref{fig:Lattice}), or that the order-disorder phase line terminates near 42 GPa and 600 K, and there is {\emph no} disordered phase above that point. Analysis of the diffraction data collected at 42 GPa and 596 K, the highest pressure at which the order-disorder transition was observed, shows that it was still clearly observed, with ordered and disordered phases easily distinguishable over a pressure difference of $\sim$2 GPa. Reducing the pressure on the sample from the highest P-T conditions reached in this study -- 47 GPa and 750 K -- would have enabled us to determine whether the order-disorder transition still existed at a lower pressure at this temperature. Unfortunately, due to the limitations of the piston-cylinder pressure cells used, it was not possible to reduce the sample pressure {\it in situ}. The nature of the order-disorder transition transition above 42 GPa therefore remains unknown.

In conclusion, we have made detailed studies of the order-disorder chain melting transition in incommensurate potassium to 47 GPa. At 20.3 GPa and 295 K the guest-atom chains are already undergoing considerable 1D thermal motion within the channels of the host structure, and on heating at 20.3 GPa long-range ordering of the chains is lost at 325 K. This disordering temperature increases with pressure up to at least 42 GPa and 596 K, the maximum P-T point where it was possible to disorder the guest chains. At pressures above 42 GPa, it was not possible to disorder the guest chains, even at 750 K, the maximum temperature reached in this study.

\section*{ACKNOWLEDGEMENTS}

The work in this paper was supported by the UK Engineering and Physical Sciences Research Council (EPSRC). Parts of this research were carried out at the light source PETRA III at DESY, a member of the Helmholtz Association (HGF). We thank H. Wilhelm, A. Kleppe and D. Daisenberger (I15, Diamond Light Source), and M. Hanfland (ID09a, ESRF) for setting up the respective beamlines. We thank S. MacLeod at AWE for use of diamond anvil cells and a vacuum vessel for resistive heating, and I. Loa at the University of Edinburgh for use of analysis software. M.I.M. acknowledges AWE Aldermaston for the award of a William Penney Fellowship, and E.E.M. and R.J.H. acknowledge financial support from EPSRC via the Scottish Doctoral Training Centre in Condensed Matter Physics.\\


\newpage

\end{document}